\documentclass[9pt,twocolumn,twoside]{osajnl}
\usepackage{eurosym}
\usepackage{exscale}
\usepackage{siunitx}
\usepackage{relsize}
\usepackage[sort&compress]{natbib}
\usepackage{graphicx,subfigure,amssymb,dcolumn,bm,amsmath}
\journal{ol}
\setboolean{shortarticle}{true}

\title{Vortex clusters bifurcating from multipoles and second-order ring solitons}
\author[1]{Hanqi Dong}
\author[2,3]{Boris A. Malomed}
\author[1*]{Zhiwei Men}

\affil[1]{Department of Physics, Jilin University, Changchun, 130012, China }
\affil[2]{Department of Physical Electronics, School of Electrical and Computer Engineering, Faculty of Engineering, Tel Aviv University, Tel Aviv 69978, Israel}
\affil[3]{Instituto de Alta Investigaci\'{o}n, Universidad de Tarapac\'{a}, Casilla 7D, Arica, Chile}
\affil[*]{Corresponding author: zwmen@jlu.edu.cn}

\ociscodes {(190.0190) Nonlinear optics; (190.6135) Spatial solitons; (070.7345) Wave propagation.}
\begin{abstract}
We address the existence, stability, and propagation dynamics of multipole solitons and vortex clusters in cubic-quintic media subject to a harmonic trapping potential. We found that vortex clusters comprising $N$ off-centered vortices with alternating topological charges  $m=\pm1$, evenly distributed on a ring, can bifurcate from a multipole soliton for $N\leq 4$ and from a second-order ring soliton for $N>4$. Rigorous linear stability analysis,  corroborated by direct numerical simulations, shows that upper-branch vortex clusters with $N=2$ and $4$ remain stable over a wide range of the propagation constant. Thus, we reveal the formation mechanism of vortex clusters.
\end{abstract}
\setboolean{displaycopyright}{true}

\begin{document}
\maketitle

The generation, propagation, and interaction of vortices in nonlinear systems have been extensively studied across various areas of physics,
including nonlinear optics \cite{YU_K_book, PROG2005_1}, Bose-Einstein condensates (BECs) \cite{Li2008, Salgueiro}, photonic cavities, and
electron beams \cite{Pismen1999,malomed2022}. Optical vortices are singular structures carrying nonzero angular momentum, characterized by a phase
circulation \cite{PROG2005_1,Falsi}. They share many properties with vortices observed in other systems, such as superfluids and BECs \cite{AS,Macri}. Vortex solitons find diverse applications, including optical trapping, microscopy, and quantum
information \cite{Torres2011}.

Beyond isolated vortices, the interplay of the vorticity and nonlinear self-trapping gives rise to composite structures. Among them, ring-shaped
vortex clusters are of special interest \cite{Liangwei2}. These complexes have been predicted in nonlinear optics and two-component BECs, thereby offering
promising avenues for exploring complex soliton configurations.

In the two-dimensional (2D) geometry, the Kerr self-focusing leads to the critical wave collapse \cite{malomed2022,PhysRevLett.133.183804}. The collapse can be suppressed by competing nonlinearities \cite{Quiroga-Teixeiro97,michinel2004square}, which arise in various physical contexts, including plasma physics \cite{zakharov1971behavior}, Bose superfluids \cite{PhysRevLett.78.1215}, and quantum droplets \cite{PhysRevLett.115.155302}. The dielectric response of some nonlinear optical materials is accurately described by combined cubic-quintic nonlinearities \cite{Lawrence:98,Cid}.

In addition to vortices, vortex-antivortex (VAV) pairs (alias vortex dipoles) and quadrupoles, nested in a localized optical beam, have been
reported too \cite{Driben_2015, PhysRevE.95.032208}. Rotating vortex clusters represent a distinct class of vortex solitons, first predicted in $D $-dimensional media with the strength of the inhomogeneous defocusing nonlinearity growing in the radial ($r$) direction faster than $r^{D}$ \cite{Kartashov:17}. In nonlocal nonlinear media, vortex pairs \cite{Shen2012} , rotating twin-vortex solitons \cite{Ye:10} and vortex breathers \cite{CHEN201969} have been studied. Nonrotating and rotating vortex clusters and lattices have been revealed in diverse settings \cite{PhysRevLett.123.160405,PhysRevA.108.053309}.

Despite the above progress, the formation and stabilization mechanism of vortex clusters is not yet fully understood. In this Letter, we report
various families of 2D multipole solitons and vortex clusters, maintained by the interplay between a harmonic-oscillator (HO) potential and cubic-quintic
(CQ) nonlinearity. Specifically, we find that clusters of $N$ vortices, with $N\leq 4$, bifurcate from upper-branch multipole solitons, or, for $N>4$,
from second-order ring solitons, with the upper branch referring to the dependence of the soliton's power on its propagation constant.

The propagation of optical beams along the $z$-axis in a bulk medium with the CQ nonlinearity and HO potential, $V(r)=\omega ^{2}r^{2}/2$, is governed
by the scaled nonlinear Schr\"{o}dinger equation for the complex field amplitude:
\begin{equation}
	i\frac{\partial \Psi }{\partial z}=-\frac{1}{2}\nabla _{\perp }^{2}\Psi
	+V\Psi -|\Psi |^{2}\Psi +|\Psi |^{4}\Psi .  \label{Eq1}
\end{equation}%
Here $\nabla _{\perp }^{2}=\partial _{x}^{2}+\partial _{y}^{2}$ is the diffraction operator, $(x,y)$ are the transverse coordinates, normalized by
the input beam width, $r=\sqrt{x^{2}+y^{2}}$, $\omega $ is the trapping frequency, and $z$ is the propagation distance scaled by the diffraction
length. Equation (\ref{Eq1}) conserves the power (norm) $U=\iint |\Psi(x,y)|^{2}dxdy$, Hamiltonian $H=\iint \left[ \frac{1}{2}\left\vert \nabla
\Psi \right\vert ^{2}+V\left\vert \Psi \right\vert ^{2}-\frac{1}{2}\left\vert \Psi \right\vert ^{4}+\frac{1}{3}\left\vert \Psi \right\vert ^{6}%
\right] dxdy$, and angular momentum $M=i\iint \Psi ^{\ast }\left( y\partial_{x}\Psi -x\partial _{y}\Psi \right) dxdy$, where $\ast $ stands for the
complex conjugate. Below, we fix $\omega =0.02$, which makes it possible to produce generic results for a relatively loose
HO trap. Stationary soliton solutions are sought for by setting $\Psi(x,y,z)=\psi (x,y)\exp (ibz)$, where $b$ is the propagation constant
and $\psi (x,y)$ is the stationary amplitude satisfying the equation
\begin{equation}
	\frac{1}{2}\nabla _{\perp }^{2}\psi -b\psi -V(r)\psi +|\psi |^{2}\psi -|\psi
	|^{4}\psi =0.  \label{Eq2}
\end{equation}

In the absence of the external potential, the CQ nonlinearity maintains the well-known flat-top profile for sufficiently high power \cite{Quiroga-Teixeiro97}. The relatively weak trapping potential confines the optical field, facilitating the formation of vortex clusters. To obtain stationary solutions of this type, we use the input
\begin{equation}
	\psi \left( x,y\right) =A\exp (-r^{4}/w^{4})\sum_{k=1}^{N}\exp \left[
	(-1)^{k}i\arctan \frac{y-y_{k}}{x-x_{k}}\right] ,  \label{input}
\end{equation}%
where the even integer $N$ denotes the total number of local unitary vortices, $k$ is the vortex number in the cluster, factor $(-1)^{k}$ implies
that the cluster is built of constituents with alternating signs of the vorticity, and $(x_{k},y_{k})=[r_{0}\cos (\theta _{k}),r_{0}\sin (\theta
_{k})]$, with $\theta _{k}=2(k-1)\pi /N$, represent positions of the vortex pivots, which are placed along a ring of radius $r_{0}$. 

The stability of the cluster states against small perturbations can be analyzed by substituting a perturbed solution into Eq.~(\ref{Eq1}), in the
usual form $\Psi (x,y,z)=\left[ \psi (x,y)+f(x,y)\exp (\lambda z)\right.
\left. +g^{\ast }(x,y)\exp (\lambda ^{\ast }z)\right] \exp (ibz)$, where $f\left( x,y\right) $ and $g\left( x,y\right) $ constitute eigenmodes of
infinitesimal perturbations, and $\lambda $ is the (complex) instability growth rate. Then, the linearization of Eq.~(\ref{Eq1}) around $\psi $ leads
to an eigenvalue problem for $f$ and $g$:
\begin{equation}
	i%
	\begin{bmatrix}
		\mathcal{L}_{\text{11}} & \mathcal{L}_{\text{12}} \\
		-\mathcal{L}_{\text{12}}^{\ast } & -\mathcal{L}_{\text{11}}^{\ast }%
	\end{bmatrix}%
	\begin{bmatrix}
		f \\
		g%
	\end{bmatrix}%
	=\lambda
	\begin{bmatrix}
		f \\
		g%
	\end{bmatrix}%
	,  \label{Eq4}
\end{equation}%
with $\mathcal{L}_{\text{11}}=\frac{1}{2}\nabla _{\perp
}^{2}-V(r)-b+2\left\vert \psi \right\vert ^{2}-3|\psi |^{4}$ and $\mathcal{L}_{\text{12}}=\psi ^{2}(1-2|\psi |^{2})$. Eigenvalues $\lambda $ were calculated  by means of the Fourier collocation method \cite{yang20103}, the underlying stationary states being
stable if Re$(\lambda )=0$ for all eigenvalues.

\begin{figure}[tbph]
	\centering
	\includegraphics[width=0.42\textwidth]{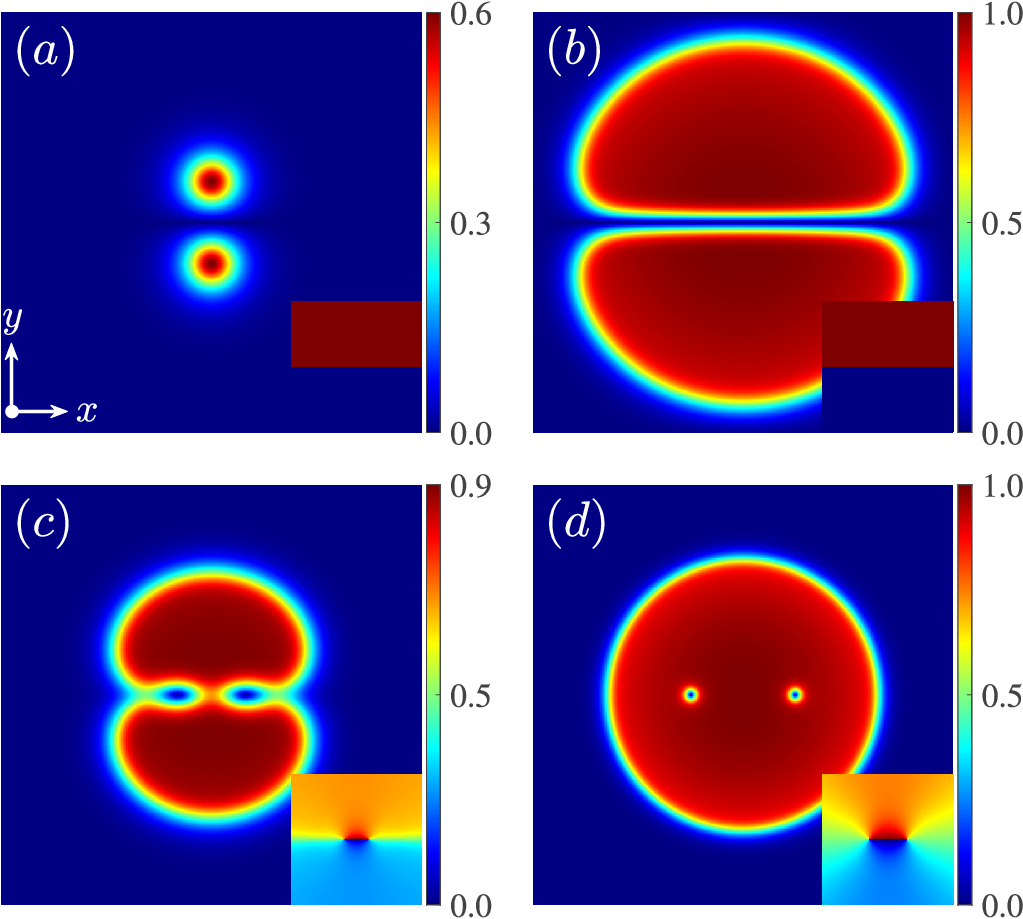} \vskip 0pc 
		\caption{The absolute value and (insets) phase distributions for lower- and upper-branch dipole solitons [(a) and (b), respectively], and for lower- and
		upper-branch VAV dipoles [(c) and (d)], which are marked in Fig. \protect\ref{fig2}(a). The propagation constants are $b=0.05$ in (a, b), $0.12$ in (c),
		and $-0.02$ in (d). Here and in other figures, the plotted domain is $|x,y|\leq 30$, except for (d), where it is $|x,y|\leq 50$.}
	\label{fig1} \vskip -0.8pc
\end{figure}

Localized nonlinear modes can bifurcate from eigenmodes of the linear version of Eq.~(\ref{Eq2}). For the 2D system with the HO potential,
the corresponding eigenvalues are $b=-(n_{x}+n_{y}+1)\omega $, where $n_{x,y} $ are the quantum numbers along the $x-$ and $y-$ directions. The
discrete eigenvalues $b=-0.02,-0.04,-0.06,-0.08,...$ (corresponding to $\omega =0.02$ and $n_{x}+n_{y}=0,1,2,3...$) form the linear spectrum.
Fundamental solitons bifurcate from the ground-state eigenmode ($n_{x}=n_{y}=0$) at $b=-0.02$, while multipole solitons bifurcate from
higher-order eigenmodes at $b=-0.04,-0.06,-0.08...$. A linear superposition of mutually degenerate orthogonal eigenmodes also yields an admitted linear
state. 


Typical examples of \textit{dipole solitons}, bifurcating from the eponymous eigenmodes at $b=-0.04$, are presented in Figs.~\ref{fig1}(a,b). The soliton
lobes in Fig.~\ref{fig1}(a) are relatively small because the low-amplitude soliton is predominantly controlled by the focusing cubic term. At larger
values of $|b|$, $|\psi |_{\mathrm{max}}$ becomes larger too, making the defocusing quintic term the dominant one in the core region of the solitons.
In the latter case, the defocusing nonlinearity induces expansion of the dipole solitons with a limited amplitude, see Fig. \ref{fig1}(b). Naturally, the two lobes in the dipole solitons are out-of-phase (having opposite signs), as shown in the insets of Figs.~\ref{fig1}(a,b).

The main finding of this work is that \textit{vortex clusters} bifurcate from multipole solitons, when the defocusing nonlinearity is dominant. For
example, at a specific value of $b$, two phase singularities with charges $m_{1,2}=\pm 1$ enter the gap between the lobes of the
vertical dipole soliton from $x=\pm \infty $. Subsequently, the lobes merge and enclose the VAV pair, as seen in Figs.~\ref{fig1}(c,d). The phase distribution presented in the insets of Figs.~\ref{fig1}(c,d) corroborate the presence of the VAV pair.

\begin{figure}[tbph]
	\centering
	\includegraphics[width=0.45\textwidth]{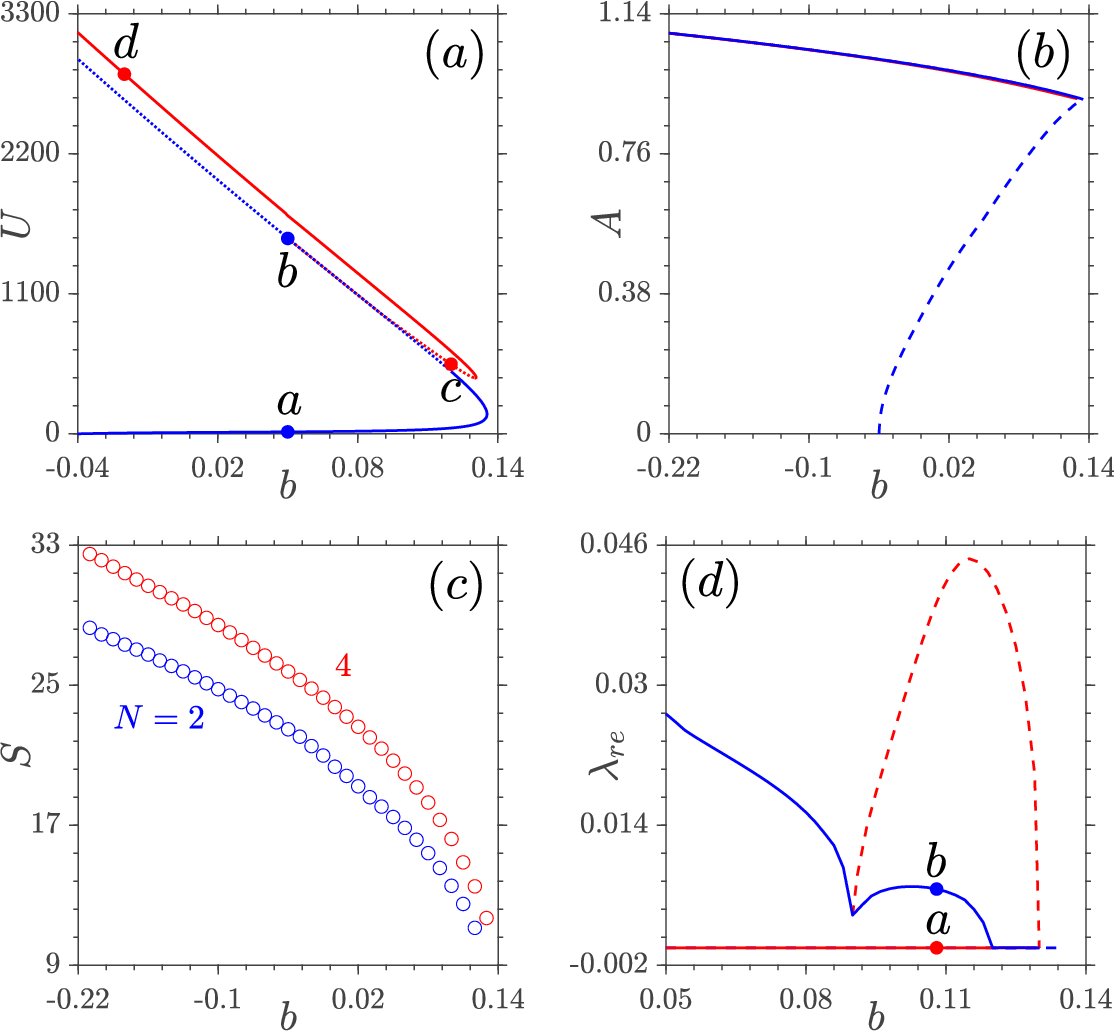} \vskip 0pc 
	\caption{(a) Power $U$ versus propagation constant $b$ for dipole solitons (blue) and VAV dipoles (red). Solid: stable; dotted: unstable. (b) The mode's amplitude $A=|\protect\psi |_{\text{max}}$ vs. $b$ for the two soliton families. (c) The geometric distance $S$ between vortex and antivortex pivot vs. $b$ for the upper-branch VAV dipoles (blue) and quadrupoles (red). (d) The maximum real part of the perturbation growth rate $\protect\lambda _{\mathrm{re}}$ vs. $b$ for dipole solitons (blue) and VAV dipoles (red). Solid and dashed lines denote upper and lower branches, respectively. Dots in (a) mark the solitons shown in Fig.~\protect\ref{fig1}. Dots in (d)
		correspond to the propagation examples in Fig.~\protect\ref{fig5}.}
	\label{fig2} \vskip -0.8pc
\end{figure}

Dipole solitons bifurcate from the eponymous linear eigenmode at point $b=-0.04$, $U=0$ [Fig.~\ref{fig2}(a)]. The power increases slowly with $b$
until reaching a turning point at $b_{\text{\textrm{turn1}}}=+0.1354$. Beyond this point, the quintic defocusing term dominates in the soliton's
core region, slowing the growth of $|\psi |_{\text{max}}$ and accelerating the expansion of the soliton, as seen in Fig.~\ref{fig1}. The gradual
transition to defocusing, combined with the confining action of the HO potential, prohibits the existence of dipole solitons for $b$ exceeding the
cutoff value (i.e., the turning point). Above the turning point, further increase of power $U$ is accommodated by the upper branch of the $U(b)$
curve in Fig.~\ref{fig2}(a). The opposite slopes of $dU/db$ clearly indicate the dominance of focusing and defocusing nonlinearities below and
above $b_{\mathrm{turn1}}$, respectively. 

VAV dipoles bifurcate from the dipole soliton at $b=0.091$. As $b$ increases further, $U$ decreases until the respective turning point $b_{\mathrm{turn2}%
}=0.1307$ is reached; afterwards, the power of the VAV dipole increases almost linearly as $b$ decreases in Fig.~\ref{fig2}(a). Note that $b_{\mathrm{turn1,2}}$ denote the power turning points of the multipole solitons and VAV clusters, respectively. VAV dipoles rapidly
expand with the increase of power $U$ [Fig.~\ref{fig1}(d)]. While $|\psi |_{\text{max}}$ quickly grows in the regime dominated by the self-focusing, the
growth nearly halts in the region of the defocusing domination, replaced by the transverse expansion of the soliton [Fig.~\ref{fig2}(b)]. The distance
between the vortex and antivortex pivots decreases monotonously with the increase of $b$ [Fig.~\ref{fig2}(c)], accompanied by shrinkage of the beam's overall size.

The results of the linear-stability analysis, based on Eqs.~(\ref{Eq4}) for the dipole solitons and VAV dipoles, are presented in Fig.~\ref{fig2}(d).
The conclusion is that upper-branch VAV dipoles are \emph{stable in their entire existence domain}. On the other hand, while the instability growth
rate of lower-branch dipole solitons vanishes, and upper-branch ones are stable in a narrow region near $b_{\text{turn1}}$, the lower-branch VAV
dipoles \emph{are completely unstable}.

\begin{figure}[h]
	\centering
	\includegraphics[width=0.42\textwidth]{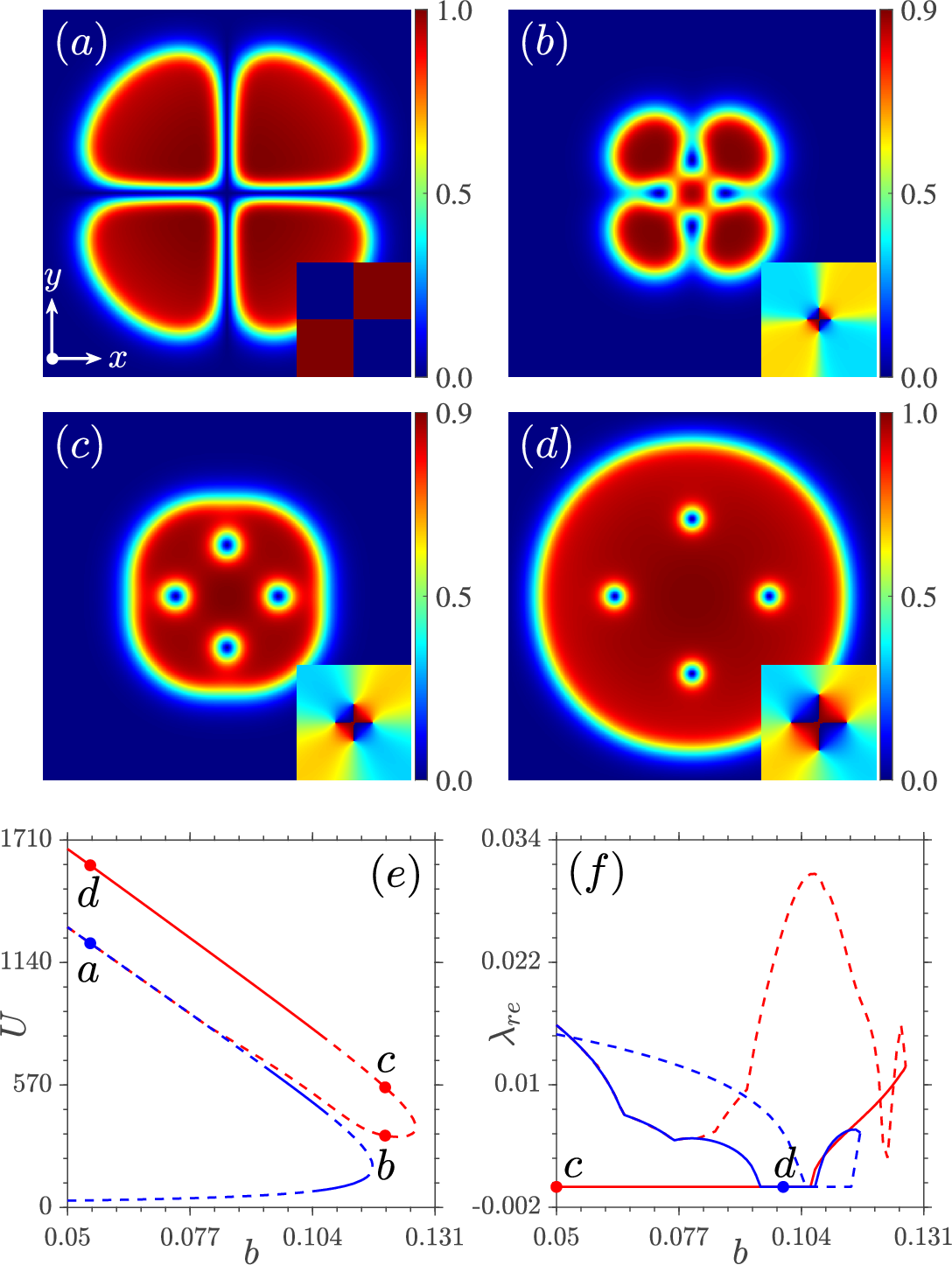} \vskip 0pc 
	\caption{The absolute value and (insets) phase distributions of upper-branch
		quadrupole solitons (a), lower-branch (b), and upper-branch (c,d) VAV
		quadrupoles marked in (e). The propagation constants are $b=0.055$ in (a, d)
		and $0.12$ in (b, c). (e) Power $U$ vs. $b$ for the
		quadrupole solitons (blue) and VAV quadrupoles (red). Solid: stable; dashed: unstable.  (f) The maximum real part of the perturbation growth rate $\protect\lambda _{\mathrm{re}}$ vs. $b$ for the quadrupole solitons
		(blue) and VAV quadrupoles (red). Solid: upper branches, dashed: lower
		branches. Dots in (f) correspond to propagation examples shown in Fig.~\protect\ref{fig5}.}
	\label{fig3} \vskip -0.8pc
\end{figure}

In addition to VAV dipoles, we also found VAV quadrupoles that bifurcate from upper-branch quadrupole solitons [Fig.~\ref{fig3}]. The quadrupole
soliton bifurcates from the linear eigenmode at $b=-0.06$ and exists up to $b_{\text{turn1}}=0.1172$, beyond which the slope of the power curve becomes
negative [Fig.~\ref{fig3}(e)]. VAV quadrupoles bifurcate from the quadrupole soliton at $b=0.081$, where four phase singularities with charges $m_{1,2,3,4}=1,-1,1,-1$ start entering gaps between four lobes. Subsequently, the four lobes merge and surround the quartet of the VAV pivots [Fig.~\ref%
{fig3}(b)]. Near $b_{\text{turn2}}=0.1267$, the envelope of a low-power VAV quadrupole exhibits a square-like shape [Fig.~\ref{fig3}(c)]. As $b$
decreases, the beam's envelope evolves into a circular shape [Fig.~\ref{fig3}(d)]. The vortex cores are more pronounced for low-power VAV quadrupoles
[Figs.~\ref{fig3}(c,d)]. The effective beam's width increases with the growth of power $U$ [Figs.~\ref{fig3}(a,d)]. At a fixed $b$, the distance
between adjacent pivots of VAV quadrupole is larger than that of the VAV dipole [Fig.~\ref{fig2}(c)] and decreases monotonously with the increase of $b$.

The stability properties of the quadrupole solitons and VAV quadrupoles clearly differ from those of their dipole counterparts [Fig.~\ref{fig3}(f)].
The lower-branch quadrupole solitons are stable in the interval $b\in \lbrack 0.105,0.115]$, whereas the upper-branch ones are stable for $b\in
\lbrack 0.095,0.107]$. Although lower-branch VAV quadrupoles are entirely unstable, their upper-branch counterparts are stable at $b<0.106$. These results indicate that the high-power VAV quadrupoles are stable modes surviving arbitrarily long propagation.

\begin{figure}[h]
	\centering
	\includegraphics[width=0.42\textwidth]{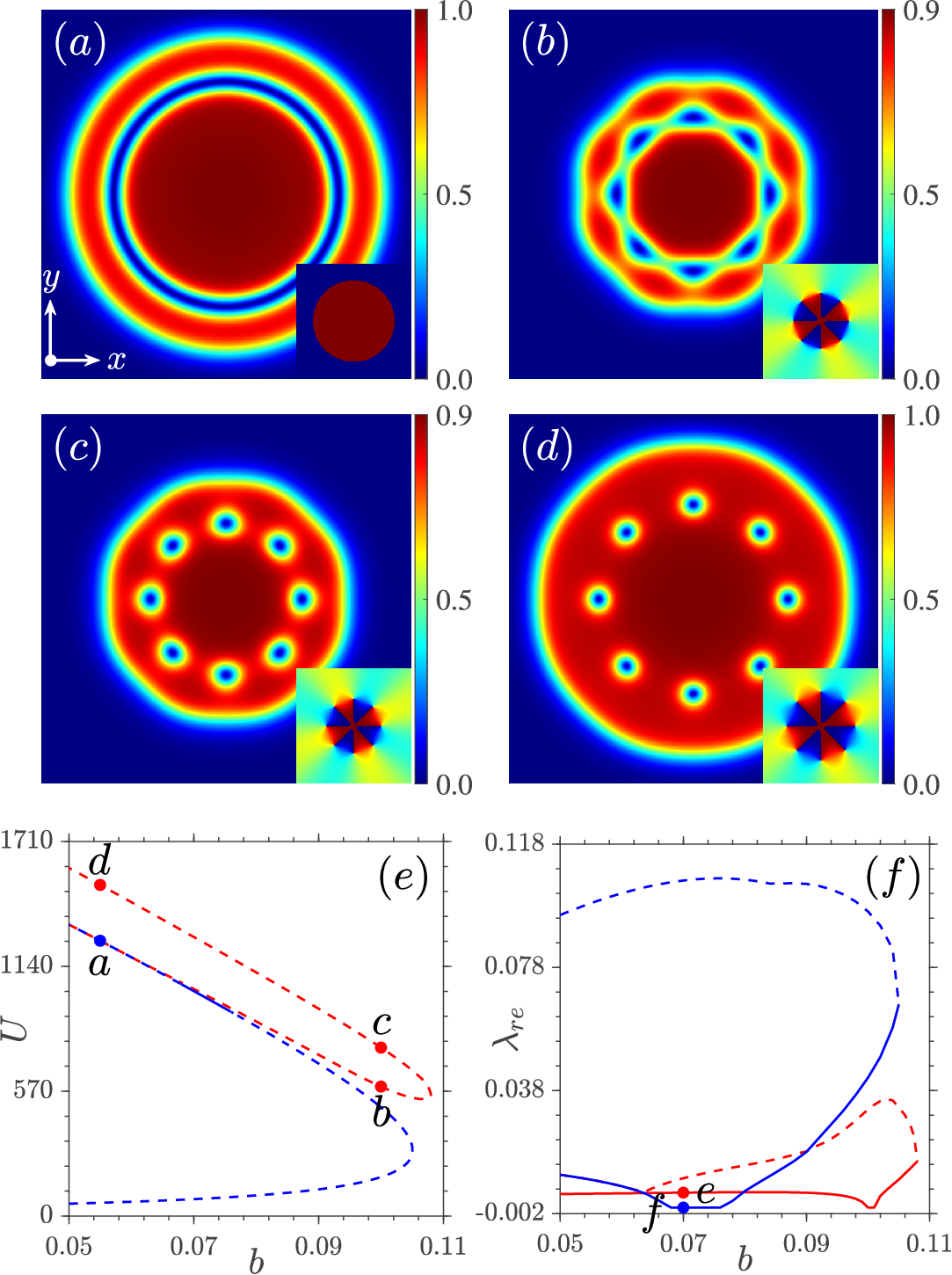} \vskip 0pc 
	\caption{The absolute value and (insets) phase distributions of the upper-branch second-order ring soliton (a), as well as lower-branch (b), and
		upper-branch (c,d) VAV octupoles marked in (e). The propagation constants are $b=0.055$ for (a, d) and $0.1$ for (b, c). (e) Power $U$ vs. $b$ for the
		second-order ring solitons (blue) and VAV ocupoles (red). Solid: stable; dashed: unstable.  (f) The maximum real part of the perturbation growth rate $\protect\lambda _{\mathrm{re}}$ vs. $b$ for the second-order ring solitons (blue) and VAV clusters with $N=8$ (red). Solid and dashed lines denote upper and lower branches, respectively. Dots in (f) correspond to the propagation examples in Fig.~\protect\ref{fig5}. }
	\label{fig4} \vskip -0.8pc
\end{figure}

Unlike the VAV dipoles and quadrupoles considered above, vortex clusters (octupoles) built of $8$ vortices and antivortices with alternating charges
bifurcate from a second-order ring soliton, which, in turn, is composed of an out-of-phase juxtaposition of an inner circular beam and an outer
ring-shaped one [Fig.~\ref{fig4}(a)]. The basic metamorphosing mechanism is that, as $b$ decreases to the bifurcation point $b=0.061$, adjacent vortices
in the clusters with a large number of elements become very close. Eventually, the attractive interaction between the neighboring vortices with
opposite charges leads to annihilation of the vortex cluster and its fusion into the second-order ring soliton.

The corresponding linear eigenmode is a superposition of mutually degenerate quadrupole modes of the linear system, which share a common eigenvalue $%
b=-0.06$. The lower-branch power in Fig.~\ref{fig4}(e) falls to zero at $b=-0.06$ (not shown here). The bifurcation point of the VAV octupole is
found at $b=0.061$, which is below the one of the VAV dipoles ($b=0.091$) and quadrupoles ($b=0.081$). Further, it is found that VAV sextupole
clusters, composed of $6$ alternating-charge vortices also bifurcate from a second-order ring soliton, at $b=0.0776$. Naturally, the envelope of the VAV
octupoles features an octagonal shape near $b_{\text{turn2}}=0.1081$ [Figs.~\ref{fig4}(b, c)]. The VAV octupoles are unstable in their nearly entire
existence domain [Fig.~\ref{fig4}(f)]. The second-order ring solitons feature a narrow stability window $b\in \lbrack 0.068,0.076]$.

\begin{figure}[htbp]
	\centering
	\includegraphics[width=0.48\textwidth]{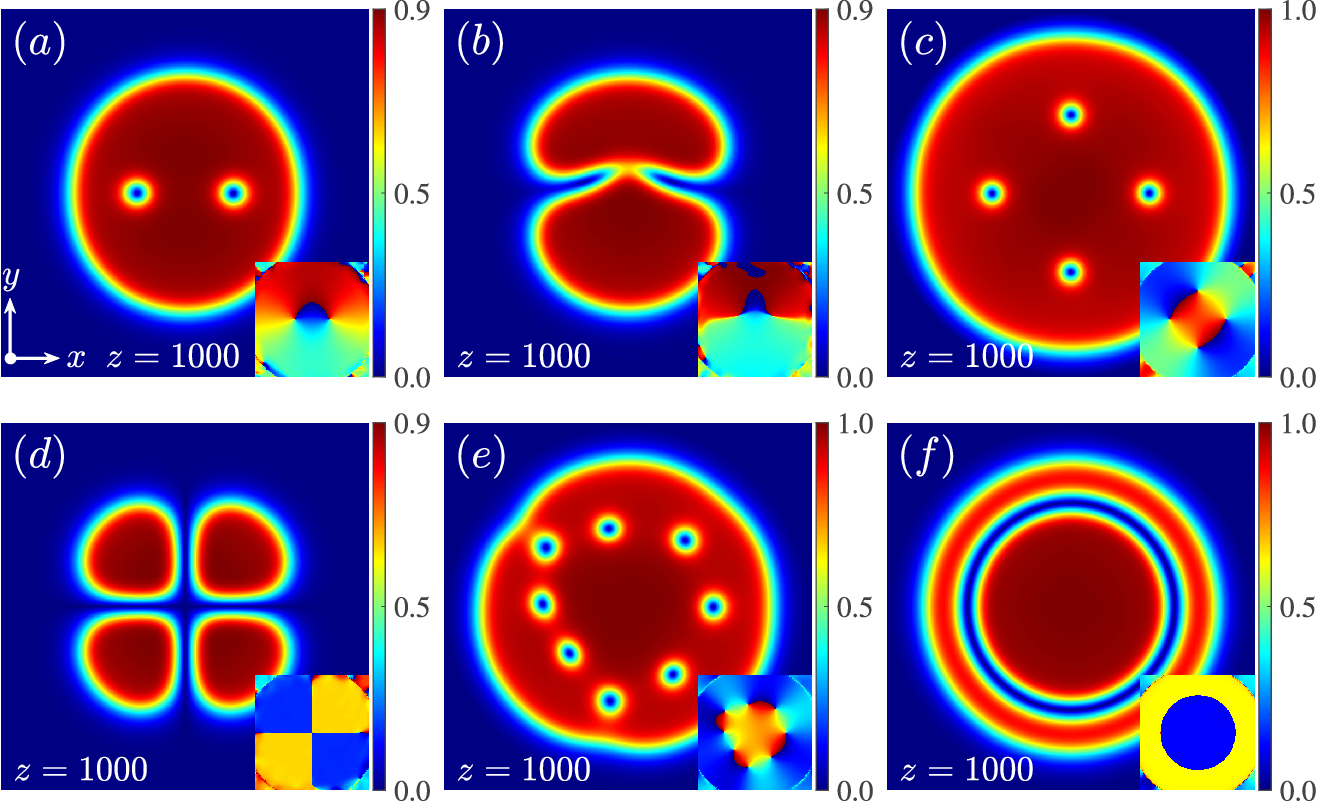} \vskip 0pc 
	\caption{The perturbed propagation of several species of stable (a,c,d,f) and unstable (b,e) 2D solitons. The absolute value and (insets) phase
		distributions of the soliton are displayed at $z=1000$. (a, b), (c,d), and (e,f) are marked by Letters in Figs.~\protect\ref{fig2}(d), \protect\ref{fig3}(f), and \protect\ref{fig4}(f), respectively. }
	\label{fig5} \vskip -0.5pc
\end{figure}

To verify the predictions of the linear-stability analysis, we performed systematic simulations of the perturbed soliton propagation by means of the
split-step Fourier method. To test the onset of possible instability, the input is taken as $\Psi (x,y,z=0)=\psi (x,y)[1+\rho (x,y)]$, where $\rho $
is random perturbation. Representative examples of stable and unstable multipole solitons, VAV clusters, and second-order ring solitons are shown
in Fig.~\ref{fig5}. Weak instability destroys the original modes after a relatively long propagation distance, as shown in Figs.~\ref{fig5}(b,e). In
contrast, stable solitons maintain their shapes intact in the course of arbitrarily long propagation in Figs.~\ref{fig5}(a,c,d,f). The predictions
of the linear-stability analysis are fully corroborated by the direct simulations in all the cases.

In conclusion, we have studied the propagation dynamics of several families of solitons in the CQ (cubic-quintic) medium confined by the HO
(harmonic-oscillator) potential. VAV (vortex-antivortex) clusters that bifurcate from the upper-branch multipole solitons are found to be stable in
a broad parameter range. The present findings can be extended to matter-wave solitons and quantum droplets trapped in HO or other circular potentials. A
challenging possibility is to generalize the study to spatiotemporal VAV clusters.

\vskip0.1pc \noindent \textbf{Disclosures.} The authors declare no conflicts of interest.
\vskip0.1pc \noindent \textbf{Data availability.} Data underlying the results presented in this paper are not publicly available at this time but may
be obtained from the authors upon reasonable request.

\vskip-3.5pc
\newpage \noindent\textbf{{\Large {References with titles}}}


\end{document}